\documentstyle[aps,prl,preprint,floats,epsfig]{revtex}

{\def\mtiny{\vrule width 0pt}
\def\mrm#1{\mathrm{#1}}
\def\DZ{\relax\ifmmode{D^0}\else{$\mrm{D}^{\mrm{0}}$}\fi}
\def\DONE{\relax\ifmmode{D_1}\else{$\mrm{D}_{\mrm{1}}$}\fi}
\def\DTWO{\relax\ifmmode{D_2}\else{$\mrm{D}_{\mrm{2}}$}\fi}
\def\KZ{\relax\ifmmode{K^0}\else{$\mrm{K}^{\mrm{0}}$}\fi}
\def\BZ{\relax\ifmmode{B^0_d}\else{$\mrm{B}^{\mrm{0}_d}$}\fi}
\def\BZp{\relax\ifmmode{B^0}\else{$\mrm{B}^{\mrm{0}}$}\fi}
\def\BZS{\relax\ifmmode{B^0_s}\else{$\mrm{B}^{\mrm{0}_s}$}\fi}
\def\DZS{\relax\ifmmode{D^{*+}}\else{$\mrm{D}^{\mrm{*+}}$}\fi}
\def\DZB{\relax\ifmmode{\overline{D}\mtiny^0}
        \else{$\overline{\mrm{D}}\mtiny^{\mrm{0}}$}\fi}
\def\KZB{\relax\ifmmode{\overline{K}\mtiny^0}
        \else{$\overline{\mrm{K}}\mtiny^{\mrm{0}}$}\fi}
\def\BZB{\relax\ifmmode{\overline{B}\mtiny^0_d}
        \else{$\overline{\mrm{B}}\mtiny^{\mrm{0}_d}$}\fi}
\def\BZBp{\relax\ifmmode{\overline{B}\mtiny^0}
        \else{$\overline{\mrm{B}}\mtiny^{\mrm{0}}$}\fi}
\def\BZBS{\relax\ifmmode{\overline{B}\mtiny^0_s}
        \else{$\overline{\mrm{B}}\mtiny^{\mrm{0}_s}$}\fi}
\def\DZC{\relax\ifmmode{\overline{D}\mtiny^0}
        \else{$\overline{\mrm{D}}\mtiny^{\mrm{0}}$}\fi}

\begin{document}

\preprint{\tighten\vbox{\hbox{\hfil CLEO CONF 99-17}
                        \hbox{\hfil}
}}

\title{Search for $\DZ\!-\!\DZB$ Mixing}

\author{CLEO Collaboration}
\date{\today}
\maketitle
\tighten

\begin{abstract} 
We report on a search for $\DZ\!-\!\DZB$ mixing made by
a study of the `\hbox{wrong-sign}' process $\DZ\!\to\!K^+\pi^-$.
The data come from 9.0~fb$^{-1}$ of integrated luminosity of $e^+e^-$
collisions at $\sqrt{s}\approx10\,$GeV,
produced by CESR and accumulated with the CLEO~II.V detector.
We measure the time-integrated rate of the `\hbox{wrong-sign}' process
$\DZ\!\to\!K^+\pi^-$, relative to that of the Cabibbo-favored
process $\DZB\!\to\!K^+\pi^-$, to be
$R_{\rm ws}=(0.34\pm0.07\pm0.06)\%$.  
By a study of that rate as a function
of the decay time of the $\DZ$, we distinguish the
rate of direct, doubly-Cabibbo-suppressed decay 
$\DZ\!\to\!K^+\pi^-$ relative to
$\DZB\!\to\!K^+\pi^-$, to be 
$R_D=(0.50^{+0.11}_{-0.12}\pm0.08)\%$.  The
amplitudes that describe
$\DZ\!-\!\DZB$ mixing, $x^\prime$ and $y^\prime$, are consistent
with zero.  The one-dimensional limits, at the 95\% C.L., that
we determine are $(1/2)x^{\prime2}<0.05\%$, and
$-5.9\%<y^\prime<0.3\%$.  All results are preliminary.
\end{abstract}
\newpage

{
\renewcommand{\thefootnote}{\fnsymbol{footnote}}

\begin{center}
M.~Artuso,$^{1}$ R.~Ayad,$^{1}$ E.~Dambasuren,$^{1}$
S.~Kopp,$^{1}$ G.~Majumder,$^{1}$ G.~C.~Moneti,$^{1}$
R.~Mountain,$^{1}$ S.~Schuh,$^{1}$ T.~Skwarnicki,$^{1}$
S.~Stone,$^{1}$
% ...
\end{center}
 
\small
\begin{center}
$^{1}${Syracuse University, Syracuse, New York 13244}\\
% ...
\end{center}

\setcounter{footnote}{0}
}
\newpage

Neutral particles such as the $\KZ$, $\DZ$, $\BZ$, and $\BZS$
mesons can evolve into their respective antiparticles, the
$\KZB$, $\DZB$, $\BZB$ and $\BZBS\,$\cite{gmp}.
Measurements of the rates and mechanisms of $\KZ\!-\!\KZB$ and
$\BZ\!-\!\BZB$ mixing have guided the form and content
of the Standard Model, and permitted
useful estimates of the 
masses of the charm and top quark masses, prior to 
direct observation of those
quarks at the high energy frontier.  

Within the framework of the Standard Model, the rate
of $\DZ\!-\!\DZB$ mixing is expected to be small, for two reasons.
First, the decays of the $\DZ$ are Cabibbo favored, while the mixing amplitudes
are doubly-Cabibbo-suppressed, and second, the GIM
cancellation\cite{gimktwz} is thought to cause substantial
additional suppression.  Many interesting extensions to the Standard
Model predict enhancements to the  rate of 
$\DZ\!-\!\DZB$ mixing\cite{hnncomp}, much as the
charm and top quarks enhance the rates of
$\KZ\!-\!\KZB$ and $\BZ\!-\!\BZB$ mixing.

We report here on a study of the process,
$\DZ\!\to\!K^+\pi^-$ (those processes obtained by the application
of charge conjugation on
all particles, such as, in this case, $\DZB\!\to\!K^-\pi^+$,
are implied throughout this report).  We use the charge of the `slow' 
pion, $\pi_s^+$, from the decay $D^{*+}\!\to\!\DZ\pi^+_s$ to deduce
production of the $\DZ$, and then we seek  the rare,
`\hbox{wrong-sign}' $K^+\pi^-$ final state, in addition to
the more frequent  `\hbox{right-sign}' final state, $K^-\pi^+$.

\begin{figure}[htpb]
\begin{center}
\epsfig{figure=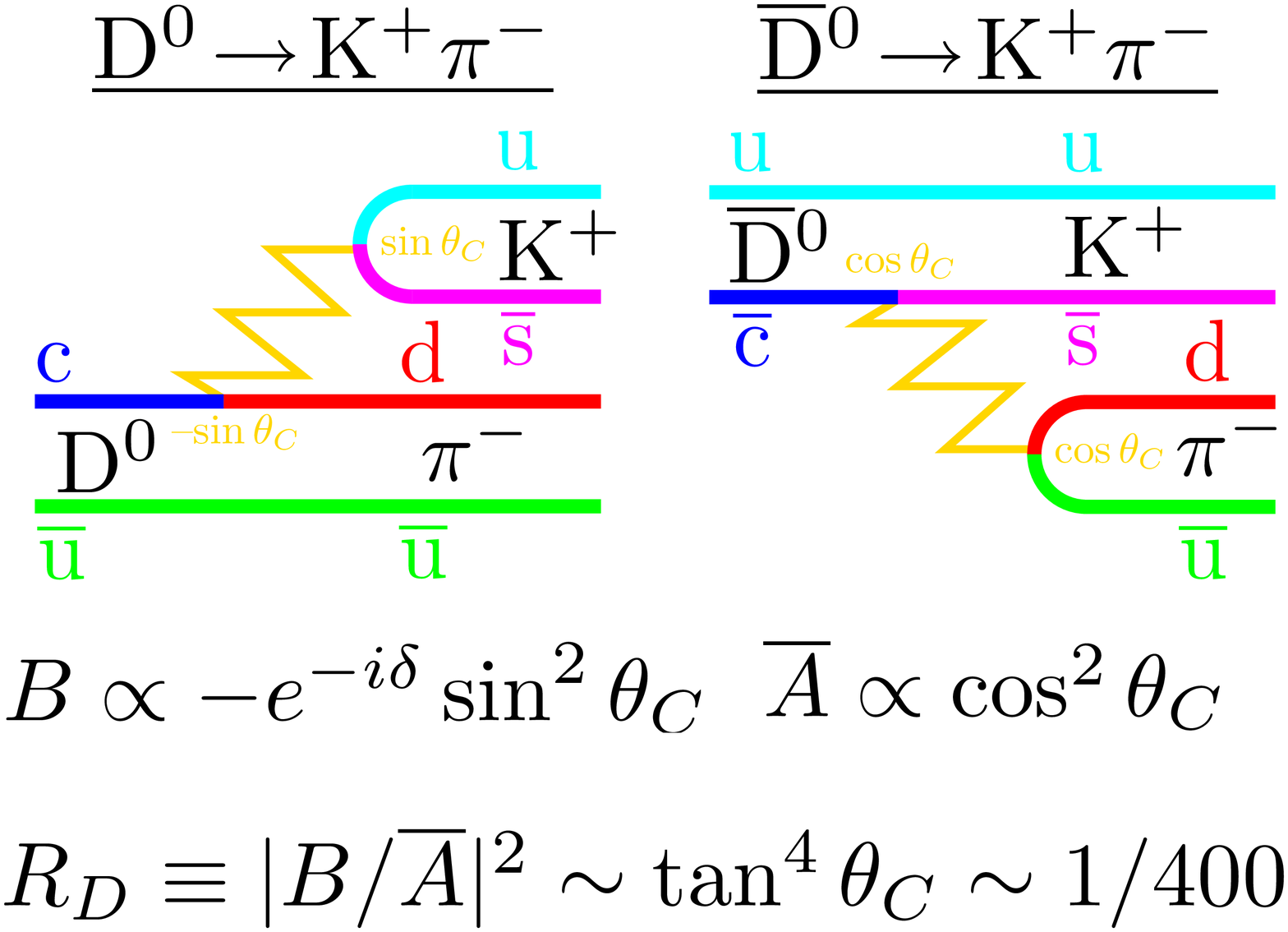,width=6.5in}
\end{center}
\caption[The Feynman Diagrams for the Decay $\DZ\!\to\!K\pi$]
{The Feynman diagrams for the decays $\DZ\!\to\!K\pi$.
The `\hbox{wrong-sign}' decay, with amplitude denoted $B$, is shown on the left, 
and the `\hbox{right-sign}' decay, with amplitude denoted $\overline{A}$,
is shown on the right.  The Cabibbo angle is $\theta_C$, and $\delta$
is a possible relative strong phase between the processes shown.
The ratio of decay rates is denoted $R_D$ and is expected to
be of order $\tan^4\theta_C\!\sim\!1/400\!=\!0.25\%$.}
\label{fig:feynkp}
\end{figure}

The \hbox{wrong-sign} process, $\DZ\!\to\!K^+\pi^-$, can proceed either
through direct, doubly-Cabibbo-suppressed decay (DCSD)
via the Feynman diagram portrayed in 
Fig.~\ref{fig:feynkp}, or through mixing followed by the Cabibbo-favored
decay (CFD), $\DZ\!\to\!\DZB\!\to\!K^+\pi^-$.  Both processes contribute
to the `\hbox{wrong-sign}' rate, $R_{\rm ws}$:
\begin{displaymath}
R_{\rm ws}=\displaystyle{\Gamma(\DZ\!\to\!K^+\pi^-)\over\Gamma(\DZB\!\to\!K^+\pi^-)}.
\end{displaymath}

\begin{figure}[htpb]
\begin{center}
\epsfig{figure=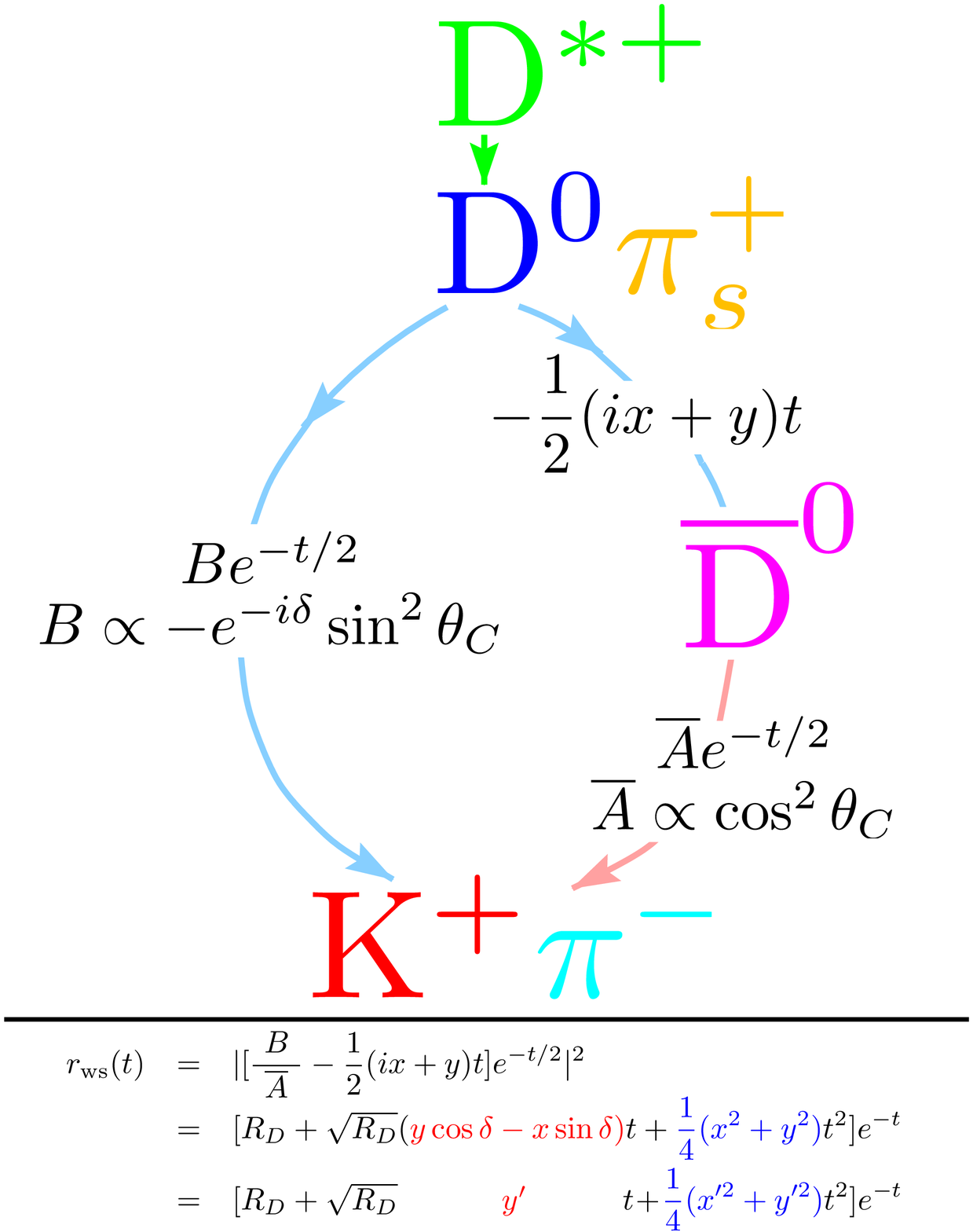,width=5.4in}
\end{center}
\caption[The Two Processes that Yield \hbox{Wrong-Sign} $K\pi$.]
{The two processes that yield \hbox{wrong-sign} $K\pi$.
The proper decay time of the $\DZ$, in units of the mean
$\DZ$ lifetime, is denoted $t$. The strong decay of
the $D^{*+}\!\to\!\DZ\pi_s^+$ specifies $t=0$, and
the charge of the slow pion $\pi_s^+$ distinguishes
a $\DZ$ from a $\DZB$.
The $\DZ$ can
undergo direct, doubly-Cabibbo-suppressed decay (DCSD), to $K^+\pi^-$,
schematically represented by the route to the left, 
with amplitude $Be^{-t/2}$.
The $\DZ$ might also mix, 
with amplitude $-(1/2)(ix+y)t$, schematically
represented by the route to the right, into
a $\DZB$, and then undergo Cabibbo-favored decay (CFD) into
$K^+\pi^-$.  The interference of these two amplitudes gives the
relative decay rate to $K^+\pi^-$ as a function of $t$, $r_{\rm ws}(t)$, shown.
The quantity $y^\prime$ is defined to be
$y\cos\delta - x\sin\delta$, and $x^\prime$ is defined analogously,
$x^\prime=x\cos\delta + y\sin\delta$.}
\label{fig:twoproc}
\end{figure}

To disentangle the two processes that could contribute to
$\DZ\!\to\!K^+\pi^-$, we study the distribution of \hbox{wrong-sign}
final states as a function of the proper decay time, $t$, 
of the $\DZ$.  The mixing amplitude grows, relative to the
decay amplitude, by a factor of $-(1/2)(ix + y)t$, as portrayed
in Fig.~\ref{fig:twoproc}.  We refer to the proper decay time $t$
in units of the mean $\DZ$ lifetime, $\tau_{\DZ}=415\pm4\,$fs\cite{RPP98}.
Then, we refer to the mixing amplitude for $\DZ\!\to\!\DZB$, in
appropriate units: those of one-half the mean $\DZ$ decay rate, 
$\Gamma_{\DZ}/2=1/[2\tau_{\DZ}]$.  The mixing amplitude through 
virtual intermediate states is $x$, and that through real 
intermediate states by $y$\cite{pt}. When conservation of
$CP$ is assumed:
\begin{displaymath}
\begin{array}{rcccl}
x&=&\displaystyle{\Delta M\over\Gamma_{\DZ}}&=&
\displaystyle{M_{12}\over\Gamma_{\DZ}/2}\vspace{2mm} \\
y&=&\displaystyle{\Delta \Gamma\over2\Gamma_{\DZ}}&=&
\displaystyle{\Gamma_{12}/2\over\Gamma_{\DZ}/2}
\end{array}.
\end{displaymath}
We use the convention that $y\!>\!0 (y\!<\!0)$ 
corresponds to a shorter(longer)
than average lifetime for eigenstates of $CP$ with the same eigenvalue
as the $\pi^+\pi^-$ state populated by $\DZ\!\to\!\pi^+\pi^-$.

It is likely, although not inevitable,
that $|x|\!\sim\!|y|$ within the Standard Model\cite{gp}.
Extensions to the Standard Model contribute to $x$ alone.

The possible interference between direct decay and mixing
would cause the number of \hbox{wrong-sign} decays, relative to 
the total number of \hbox{right-sign}
decays, as a function of $t$, $r_{\rm ws}(t)$ 
to be\cite{timev,bsn}
\begin{equation}
r_{\rm ws}(t)=[R_D + \sqrt{R_D} y^\prime t + 
         \displaystyle{1\over4}(x^{\prime2}+y^{\prime2})t^2]e^{-t},
\label{eq:rws}
\end{equation}
where $R_D$ is the ratio of \hbox{wrong-sign} to \hbox{right-sign} rates
of direct decay.
The $\DZ\!-\!\DZB$ mixing amplitudes 
$x^{\prime}$ and $y^{\prime}$ are defined by:
\begin{displaymath}
\begin{array}{rcl}
 y^\prime&\equiv&y\cos\delta\!-\!x\sin\delta\\
 x^\prime&\equiv&x\cos\delta\!+\!y\sin\delta
\end{array},
\end{displaymath}
where $\delta$ is a possible strong phase between the \hbox{wrong-sign}
and \hbox{right-sign} decay amplitudes.  In a sense, use of a hadronic
final state, such as $K\pi$, `filters' the mixing amplitudes, much as
a polarizer filters light.  There are plausible arguments
that $y'>0$\cite{bigibj}, and $\delta<13^\circ$\cite{wfsp,brpa}.

We report here on the analysis of data accumulated between 1995 and
1999 from an integrated
luminosity of $9.0\,$fb$^{-1}$ of $e^+e^-$ collisions with
$\sqrt{s}\approx10\,$GeV at the Cornell Electron Storage Rings (CESR).
The data were taken with CLEO~II multipurpose 
detector~\cite{ctwo}, which includes two cylindrical drift
chambers in a superconducting solenoid ($B=1.5\,$T) for measurement
of the three-momentum $\vec{p}$ of charged particles, a cylindrical array
of CsI crystals for measurement of the energies of photons and electrons,
and a system of iron absorbers interleaved with proportional chambers for
the identification of muons. These systems cover 
approximately 80\% of the solid angle
around the $e^+e^-$ annihilation point.

In 1995 a silicon vertex detector (SVX) was installed~\cite{csvx},
that enables both precise reconstruction of the proper lifetime of short-lived
particles such as the $\DZ$, as well as improved reconstruction of
the angle $\theta$ between the $e^+e^-$ beams and the momentum vector
of low-momentum charged particles, such as the slow charged pion from the
decay $D^{*+}\!\to\!\DZ\pi_s^+$.  Also, in 1995, the gas in the larger
CLEO~II drift chamber was changed from argon-ethane to a helium-propane
mixture, resulting in improved momentum and mass resolution, and improved
particle identification by specific ionization $(dE/dx)$.
We refer to this revised configuration of the CLEO~II detector
as  CLEO~II.V.

\begin{figure}[htpb]
\begin{center}
\epsfig{figure=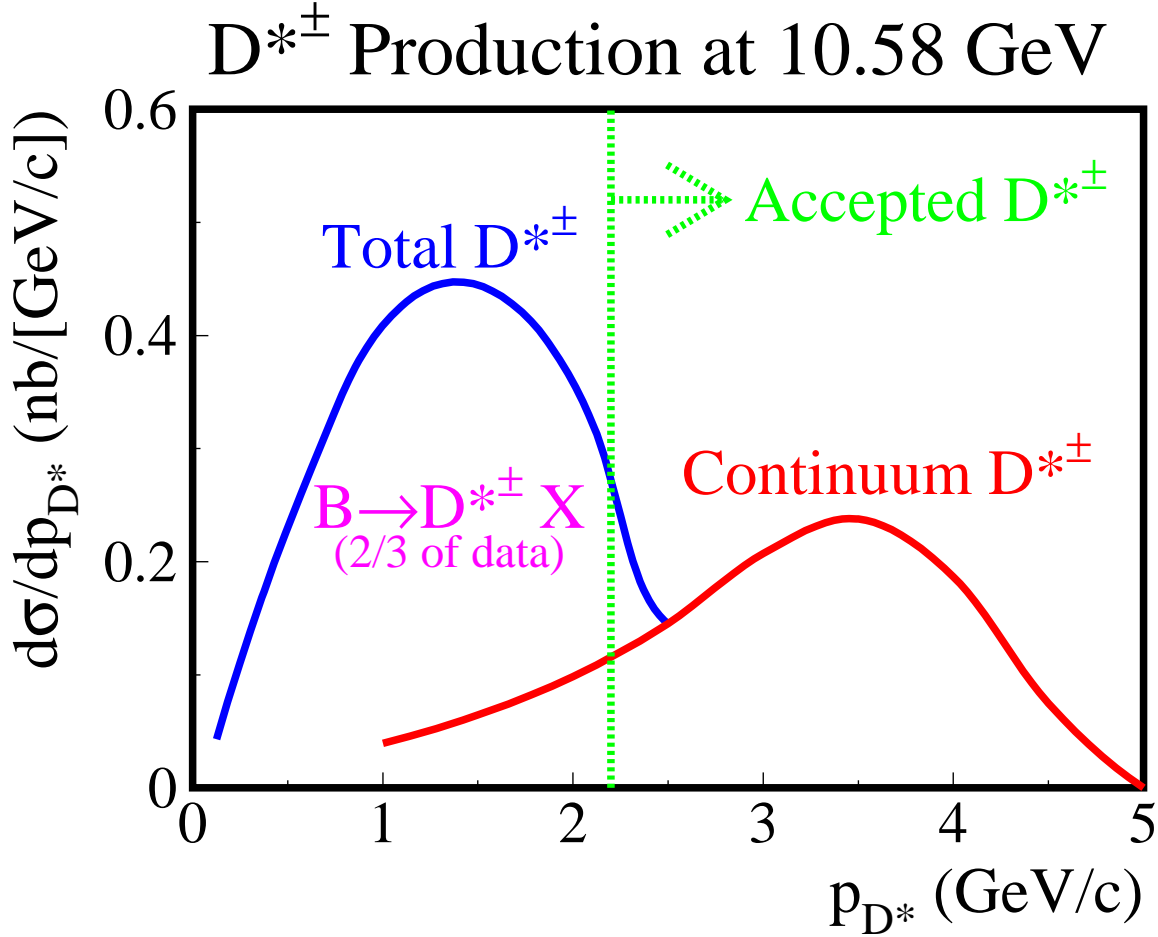,width=6.5in}
\end{center}
\caption[The $D^{*\pm}$ Production Cross Section.]
{The $D^{*\pm}$ production cross section at 
$\sqrt{s}=10.58\,$GeV\protect\cite{RPP98,moneti}.
The $D^{*\pm}$ at higher momenta
result from the continuum production:
$e^+e^-\!\to\!c\overline{c}$, followed
by hadronization of charmed quarks to a $D^{*\pm}$,
and these $D^{*\pm}$ are present for all of our data.
The component of $D^{*\pm}$ at lower momenta are present in the
2/3 of the data taken with $\sqrt{s}$ at the mass
of the $\Upsilon(4s)$, and these softer $D^{*\pm}$
result from the decay $B\!\to\!D^{*\pm}X$.  We accept
$D^{*\pm}$ with momenta greater than $2.2\,$GeV/$c$;
most of these are from continuum production.}
\label{fig:dsp}
\end{figure}

We reconstruct candidates for the decay sequences $D^{*+}\!\to\!\pi^+_s\DZ$,
followed by either $\DZ\!\to\!K^+\pi^-$ (\hbox{wrong-sign}, or WS) or
$\DZ\!\to\!K^-\pi^+$ (\hbox{right-sign}, or RS). The sign of the slow charged
pion, either $\pi^+_s$ or $\pi^-_s$, identifies (`tags') the
charm state at production (`$t=0$') as either $\DZ$ or
$\DZB$.  The broad features of the
reconstruction are similar to those employed in the recent
CLEO measurement of the $D$ meson lifetimes~\cite{dlife},
but there are four principal differences.  First, we
accept candidates with total momentum, $p_{D^*}$, as low
as $2.2\,$GeV; we show the approximate $D^*$ 
production cross section as a function of $p_{D^*}$ in
Fig.~\ref{fig:dsp}. Second, we fully exploit the three-dimensional
tracking capability of the SVX.  We reject candidates where the
daughter tracks form poor vertices in three-dimensional space,
and we sharpen the reconstructed direction of the $\pi^+_s$ momentum,
thereby improving our resolution for reconstruction of $Q$, the
(small) energy released in the $D^{*+}\!\to\!\pi^+_s\DZ$ decay.
We denote by $M$ the reconstructed mass of the two daughters of the
$\DZ$, and by $M_{K\pi\pi}$ the reconstructed mass of all three particles
in the candidate.  We reconstruct $Q=(M_{K\pi\pi}-M-m_{\pi})c^2$, where
$m_{\pi}$ is the mass of a charged pion.
Third, we require candidates to be well-measured in
$Q$, and in $M$.
Fourth, we require candidates to pass two `kinematic' requirements,
designed to suppress backgrounds from $\DZ\!\to\!\pi^+\pi^-$,
$\DZ\!\to\!K^+K^-$, $\DZ\!\to\, >\!2\,$bodies, and from cross-feed
between WS and RS.  For the first kinematic
requirement, for $\DZ\!\to K^+\pi^-$ candidates we evaluate 
the mass $M$ under the
three alternate hypotheses $\DZ\!\to\!\pi^+\pi^-$,
$\DZ\!\to\!K^+K^-$, and $\DZ\!\to\!\pi^+K^-$.  If any one of the
three masses so computed falls within
$30\,$MeV/c$^2$ (approximately $4.5\,\sigma$)
of the $\DZ$ mass, the $\DZ\!\to K^+\pi^-$  candidate is rejected.  
A conjugate requirement is made for the RS decays.
The second kinematic requirement rejects one
of the two configurations of very asymmetric decay, with a requirement
that the decay angle $\theta^*$ of the hypothesized \emph{kaon}, $\theta^*$,
in the $\DZ$ rest frame, with respect to the $\DZ$ boost, satisfy
$\cos\theta^*<0.8$.  This requirement removes candidates with slow
pions from the $\DZ$ decay.
 
The CLEO drift chamber system allows reconstruction of the
specific ionization ($dE/dx$) deposited by the passage
of charged particles, permitting an independent assessment
of the identification of a charged particle as either a $\pi$ or a $K$.
We require that the $dE/dx$ be within three standard deviations
($\sigma$) of the hypothesis used in the reconstruction 
of $M$ and $Q$; this is
a loose, consistency requirement.  We tighten $dE/dx$ criteria
only for the evaluation of systematic errors.

\begin{figure}[htpb]
\begin{center}
\epsfig{figure=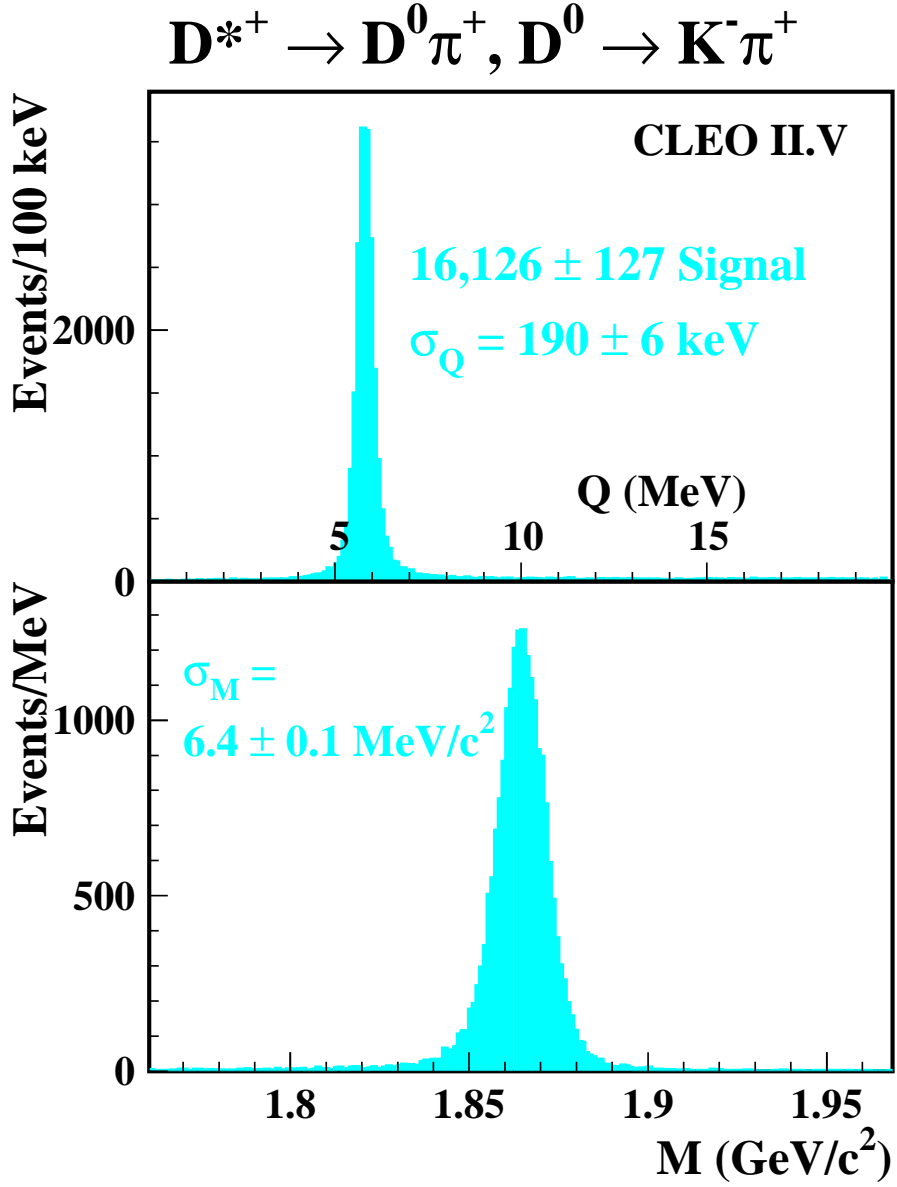,width=5.30in}
\end{center}
\caption[Signal for the \hbox{Right-Sign} Process $\DZ\!\to\!K^-\pi^+$]
{Signal for the \hbox{right-sign} process $\DZ\!\to\!K^-\pi^+$.
The events in the top and bottom plots come from the same sample, and
are all candidates for
the decay $D^{*+}\!\to\!\pi^+_s\DZ$ followed by
$\DZ\!\to\!K^-\pi^+$.
In the top plot, the
horizontal axis is $Q$, the energy released in the decay
of the $D^{*+}$, for events where the reconstructed mass $M$
of the $K\pi$ system falls within 15 MeV/c$^2$ of the $\DZ$ mass.
In the bottom plot, the
horizontal axis is $M$,
for events where $Q$
falls within 450 keV/c$^2$ of its nominal value.  The total
number of \hbox{right-sign} signal events is $16126\pm127$.
The experimental resolution function in $Q$ is non-gaussian,
due to a substantial dependence of the width of the 
specific (gaussian) resolution on the particular decay configuration; the
$\sigma_Q$ shown is the core value from a fit to a bifurcated Student's $t$
distribution.  The result for $\sigma_M$ comes from a fit to a double
gaussian.}
\label{fig:qmrs}
\end{figure}

A total of $16126\pm127$ candidates for the \hbox{right-sign} decay
pass all requirements; their distribution in $Q$ and $M$ is
shown in Fig.~\ref{fig:qmrs}.

\begin{figure}[htpb]
\begin{center}
\epsfig{figure=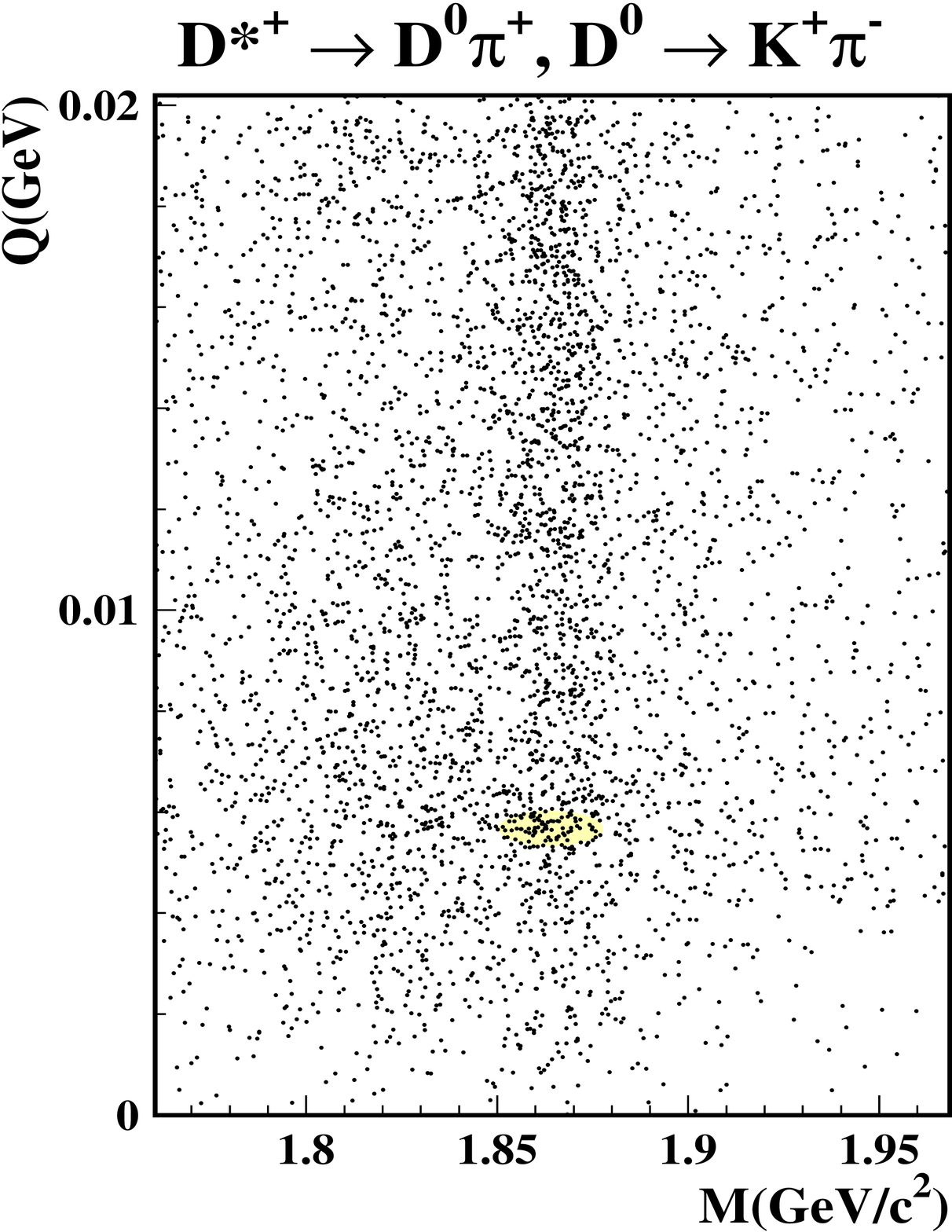,width=5.4in}
\end{center}\caption[Wrong Sign Scatter Plot]
{Scatter plot of $Q$ versus $M$ for the candidates
for $\DZ\!\to\!K^+\pi^-$.  The signal region is lightly shaded.
The most prominent feature is the vertical `ridge' at
$M_{\DZ}$ ; this results from the combination of
the CFD $\DZB\!\to\!K^+\pi^-$ with a random slow pion $\pi^+$,
which combine to fake a $D^{*+}$.
The pervasive, smooth background consists of random combinations
of charged tracks, referred to as $c\overline{c}$ and $uds$ background
in the text.
The background density for $M\!<\!M_{\DZ}$ exceeds
that for $M\!>\!M_{\DZ}$ simply due to phase space
for the random combinations, and also due to true $\DZ$ decays where a
particle has been missed, resulting in a reduction in $M$.
The so-called `PV' background, described in the text, causes an increase
in density near the signal in $Q$, but at $M<M_{\DZ}$.}
\label{fig:wi}
\end{figure}

The population of the \hbox{wrong-sign} candidates is shown on a scatter plot
of $Q$ versus $M$ in Fig.~\ref{fig:wi}.  In Fig.~\ref{fig:wi},
several prominent background features are evident.  The combination
of $\DZB\!\to\!K^+\pi^-$ with random slow pions, $\pi^+$,  causes the vertical
`ridge,' with $M$ equal to the mass of the $\DZ$, $M_{\DZ}$.
The phenomenon is sometimes called `dilution', and we
will refer to it as  `random $\pi^{\pm}+\DZ/\DZB$'.  A diffuse background
is also evident, due to the random combination of charged particles
from $e^+e^-\!\to$light quarks, and
from $e^+e^-\!\to\!c\overline{c}$, which we will refer to as
combinatoric `$uds$' and `$c\overline{c}$,' respectively.
There is also a broad enhancement near the signal
in $Q$, but at $M<M_{\DZ}$, due to processes such as
$\DZ\!\to\!K^-\rho^+$, followed by a very asymmetric decay 
$\rho^{+}\!\to\!\pi^0\pi^+$, resulting in a $\pi^0$ that is
nearly at rest.  Since a number of
modes that we model can produce such behavior,
including $\DZ\!\to\!\rho^+\pi^-$, $\DZ\!\to\!\rho^-\pi^+$,
and $\DZ\!\to\!K^{*-}\pi^+$, we
refer to this type of background as `PV',
for pseudoscalar-vector.

\begin{figure}[htpb]
\begin{center}
\epsfig{figure=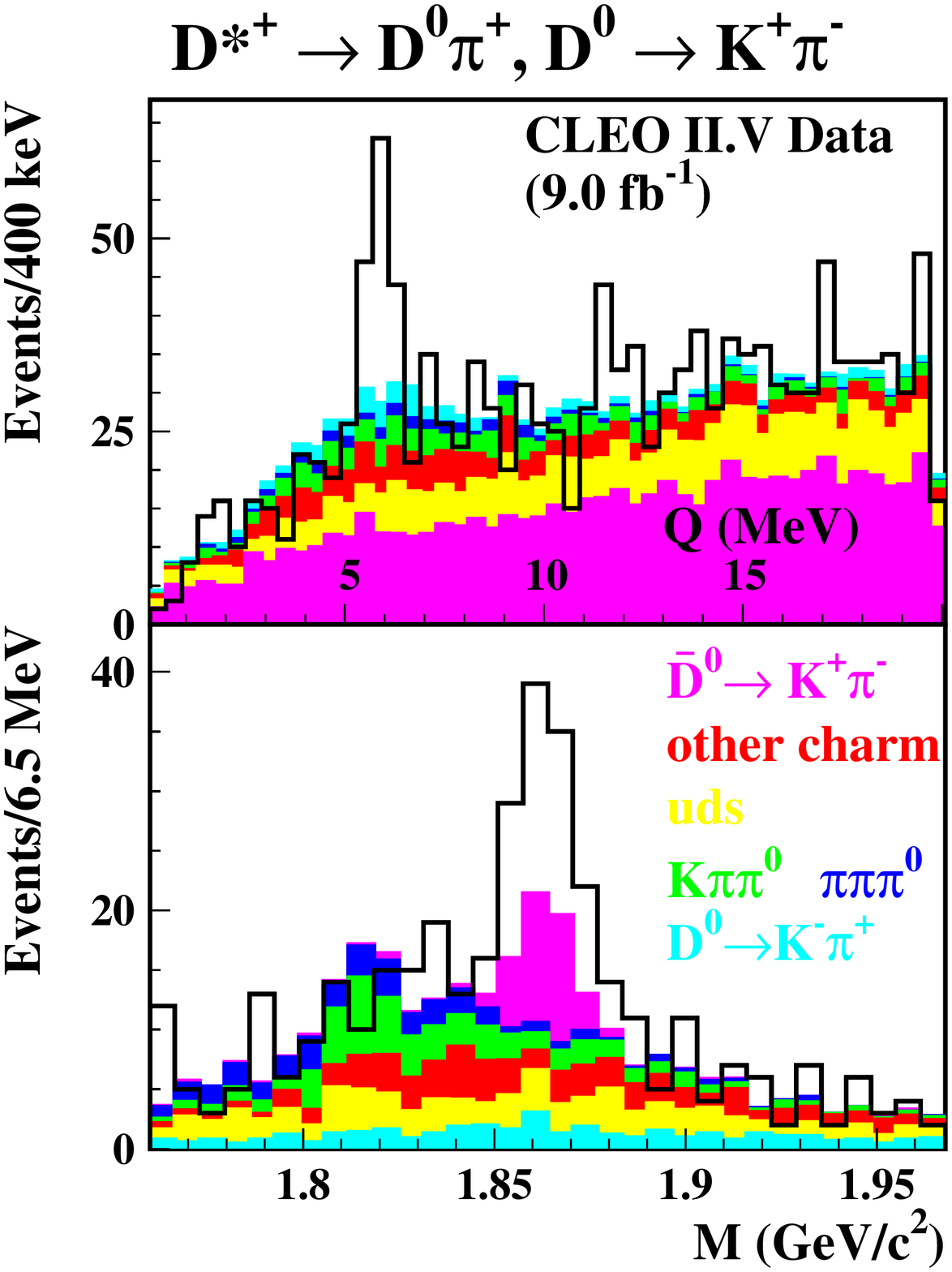,width=5.5in}
\end{center}
\caption[Signal for the \hbox{Wrong-Sign} Process $\DZ\!\to\!K^+\pi^-$]
{Signal for the \hbox{wrong-sign} process $\DZ\!\to\!K^+\pi^-$.
See also the caption for the \hbox{right-sign} analog, Fig.~\ref{fig:qmrs}.
The solid lines are the data, and the colored regions are the
contributions from various background sources from the fit,
as labeled, with shapes from simulation, 
and levels from a fit to the two-dimensional
plane of $Q$ and $M$.  The fit to the signal peak is not shown.
The results of the fit are summarized in
Table~I.}
\label{fig:qmws}
\end{figure}

To deduce the \hbox{wrong-sign} rate, $R_{\rm ws}$, we perform a 2-dimensional
fit to the region of the $Q$ versus $M$ plane shown in Fig.~\ref{fig:wi}. 
Each of the background shapes: `random $\pi^{\pm}+\DZ/\DZB$', 
$c\overline{c}$, $uds$, and PV is taken from
Monte Carlo simulated data, which statistics corresponding
to 90~fb$^{-1}$ of integrated luminosity,
but the normalization of each component 
is allowed to float in the fit; only the $uds$ shows a significant
difference with its expected contribution; the fit exceeds expectation
by approximately a factor of two.
The \hbox{wrong-sign} signal shape is taken directly from a seven $\sigma$ 
ellipse around the \hbox{right-sign} signal.  The fitted background
composition, superposed on the data, and projected on to $Q$ and $M$ are shown
in Fig.~\ref{fig:qmws}.  The event yields in the signal region from
the fit are summarized in Table~I.

\begin{center}
\begin{minipage}[t]{5.5in}
TABLE I. Event yields in a signal region of $2.4\,\sigma$
centered on the nominal $Q$ and $M$ values,
for the various  categories of signal and background.  The
total number of candidates is 107.  The first five
rows are from the
fit, in the two dimensions of $Q$ and $M$, to the data
shown in Fig.~\ref{fig:wi}.  Projections of the background
components of the fit are shown in
Fig.~\ref{fig:qmws}.  
The errors are statistical alone,
and for the background components, are the errors on the \emph{mean}
yield in the signal region.  The sixth line results from a fit
to the data \hbox{right-sign} data in Fig.~\ref{fig:qmrs}.
\begin{center}
\begin{tabular}{cc}
\textbf{Component}            & \textbf{\# Events} \\ \hline
$\DZ\!\to\!K^+\pi^-$ (WS Signal) & $54.8\pm10.8$ \\
random $\pi^{\pm}+\DZ/\DZB$   & $24.3\pm1.8$ \\
$c\overline{c}$               & $12.3\pm0.8$ \\
$uds$                         & $8.6\pm0.4$ \\ 
PV                            & $7.0\pm0.4$ \\ \hline
$\DZB\!\to\!K^+\pi^-$ (RS Normalization)\hspace{0.5cm} & $16126\pm126$ \\ \hline
\end{tabular}
\end{center}
\end{minipage}
\end{center}

No acceptance corrections are need to directly compute,
from Table~I, $R_{\rm ws}=(0.34\pm0.07)\%$.  The dominant
systematic errors all stem from the potentially inaccurate modeling
of the initial and acceptance-corrected shapes of the background
contributions in the $Q$-$M$ plane.  We assess these systematic
errors by substantial variation of the fit regions, $dE/dx$ criteria,
and kinematic criteria; the total systematic error we assess is
$0.06\%$.  

Our complete result for $R_{\rm ws}$ is summarized in Table~II.

There are two directly comparable measurements of $R_{\rm ws}$: one
is from CLEO~II\cite{tl}, $R_{\rm ws}=(0.77\pm0.25\pm0.25)\%$ which used
a data set independent of that used here; the second is
from Aleph\cite{alep},
$R_{\rm ws}=(1.84\pm0.59\pm0.34)\%$; comparison of our result and these
are marginally consistent with $\chi^2=6.0$ for 2 DoF, for a C.L. of 5.0\%.

\begin{center}
\begin{minipage}[t]{5.5in}
TABLE II. Result for $R_{\rm ws}$.  For the branching ratio
${\cal B}(\DZ\!\to\!K^+\pi^-)$, we take the absolute branching
ratio ${\cal B}(\DZB\!\to\!K^+\pi^-)=(3.85\pm0.09)\%$, and the
third error results from the uncertainty in this absolute branching ratio.
\begin{center}
\begin{tabular}{cl}
\textbf{Quantity}            & \textbf{Result} \\ \hline
$R_{\rm ws}$ & $(0.34\pm0.07\pm0.06)\%$ \\
$R_{\rm ws}/\tan^4\theta_C$ & $(1.28\pm0.25\pm0.21)$ \\
${\cal B}(\DZ\!\to\!K^+\pi^-)$\hspace{0.5cm} & $(1.31\pm0.26\pm0.22\pm0.03)\!\times\!10^{-4}$ \\ \hline
\end{tabular}
\end{center}
\end{minipage}
\end{center}

We have split our sample into candidates for $\DZ\!\to\!K^+\pi^-$
and $\DZB\!\to\!K^-\pi^+$.  There is no evidence for a $CP$-violating
time-integrated asymmetry.  From Table~II, it is straightforward 
to evaluate the $1\sigma$ statistical error on the $CP$ violating 
time-integrated asymmetry as $\sqrt{107}/54.8=19\%$.

Given the absence of a significant time-integrated $CP$ asymmetry,
we undertake a study of the decay time dependence of the
\hbox{wrong-sign} rate in which $CP$ conservation is assumed.  Our
fits use Equation~\ref{eq:rws}, which describes the \hbox{wrong-sign}
decay time dependence, \emph{include} the term that is linear in $t$,
unless specifically noted.
The fit variable $y^\prime$ is allowed to vary over \emph{all} real values, 
and we thereby account for the principal objection\cite{E791} to 
an earlier analysis of $\DZ\!-\!\DZB$ mixing\cite{E691}.  We believe
that the fact that our acceptance in $t$ extends all the way to
zero lifetimes, while the acceptance for the fixed target experiments E691 and
E791\cite{E691,E791} dropped  near $t\!\sim\!1/2$, prevents the
loss in sensitivity that E791\cite{E791} reported.

\begin{figure}[htpb]
\begin{center}
\epsfig{figure=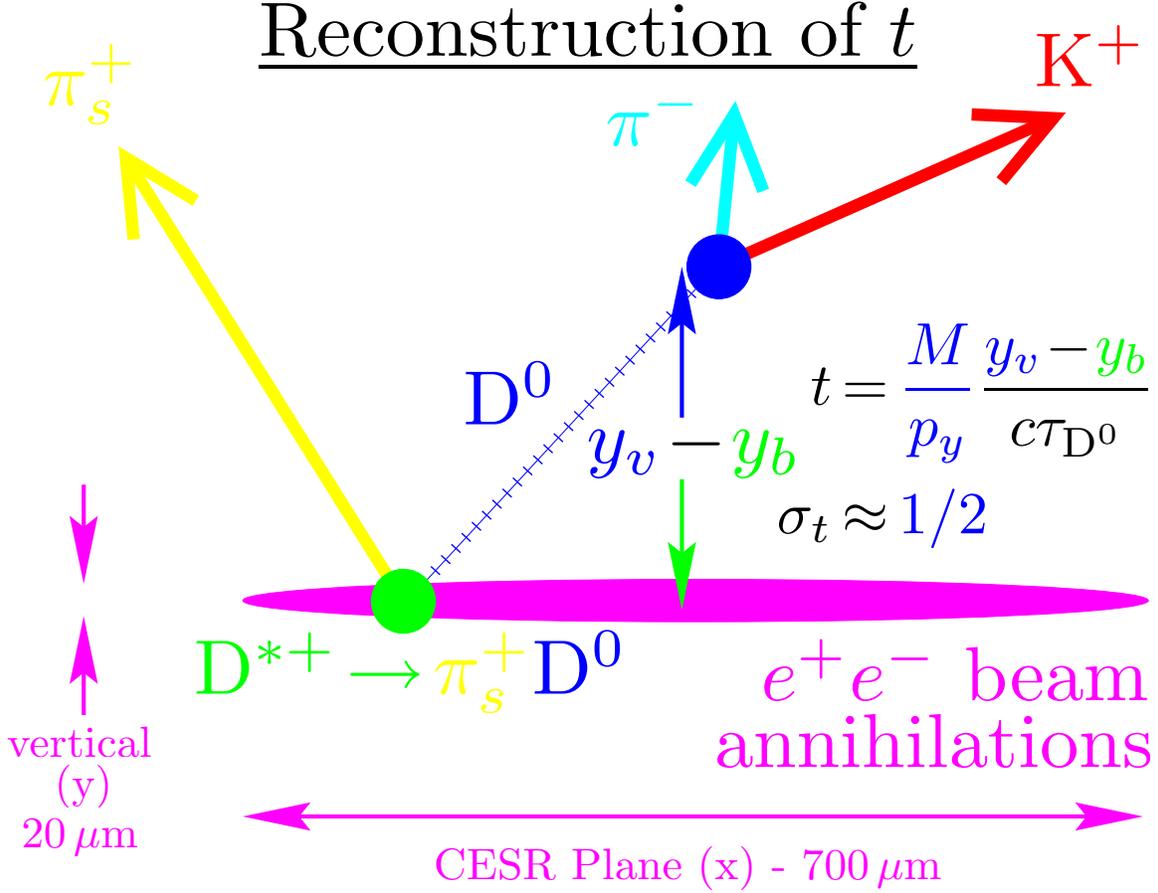,width=6.0in}
\end{center}
\caption[The Reconstruction of $t$.]{
The reconstruction of $t$.  We measure the proper decay time of the
$\DZ$, $t$, in multiples of the mean $\DZ$ lifetime,
or $t=\tau/\tau_{\DZ}$, where $\tau$ is the proper decay time
measured in seconds.  As shown in this diagram,
the region of $e^+e^-$ annihilation is very small in the vertical,
or $y$, where the region extends only $20\,\mu$m at FWHM.  In the 
bend plane of the CESR collider, or $x$, 
the FWHM extends $700\,\mu$m,
and along the beam axis or $z$, the FWHM
is $30,000\,\mu$m.  The typical three-dimensional
flight distance of the $\DZ$
is $180\,\mu$m, which is small compared to the $x$ and $z$ extents
of the luminous region.  We therefore reconstruct $t$ 
using the displacement 
of the $\DZ$ decay point, or vertex, in $y$ from the beam centroid
using the formula as shown.  The symbol $y_v$ refers to the 
$K^+\pi^-$ vertex, and $y_b$ refers to the beam centroid.  
The typical resolution in $t$
is $\sigma_t\approx1/2$, and $\sigma_t$ becomes large 
when the direction of the $\DZ$ flight is near to the
horizontal.  The error in extrapolation of the $\pi^+_s$ back
to the luminous region is large, due to multiple scattering in
the beam pipe, and indeed, reconstruction of the $\pi^+_s$ direction
benefits greatly by constraint of the $\pi_s^+$ to the intersection
of the beam and the extrapolated $\DZ$ flight path.  Use of other tracks
in the event typically do not improve reconstruction of the estimate of
the $D^{*+}$ production point beyond that shown.}
\label{fig:rect}
\end{figure}

We reconstruct $t$ using only the vertical, or $y$, component of the
flight distance of the $\DZ$.  This reconstruction is guided by
the physical dimensions of the CESR luminous region\cite{dynb}
and is portrayed in Fig.~\ref{fig:rect}.  We reconstruct the
$\DZ$ decay point, $(x_v,y_v,z_v)$, with a resolution that is
typically $40\,\mu$m in each dimension.   We measure the
centroid of the luminous region, $(x_b,y_b,z_b)$ with suitable
hadronic events in blocks of data that typically contain integrated
luminosities of several pb$^{-1}$.  Runs where $y_b$ is poorly
measured are discarded, and the dominant error on $y_v-y_b$ comes from
$y_v$.  We reconstruct $t$ as:
\begin{displaymath}
t=\displaystyle{M\over p_y}\times
\displaystyle{y_v-y_b\over c\tau_{\DZ}}
\end{displaymath}
where $p_y$ is the $y$-component of the total momentum of the
$K^+\pi^-$ system.  The error $\sigma_t$ is typically $1/2$,
although when the $\DZ$ direction is near the horizontal, $\sigma_t$
can be large; we reject candidates with $\sigma_t>3/2$.

\begin{figure}[htpb]
\begin{center}
\epsfig{figure=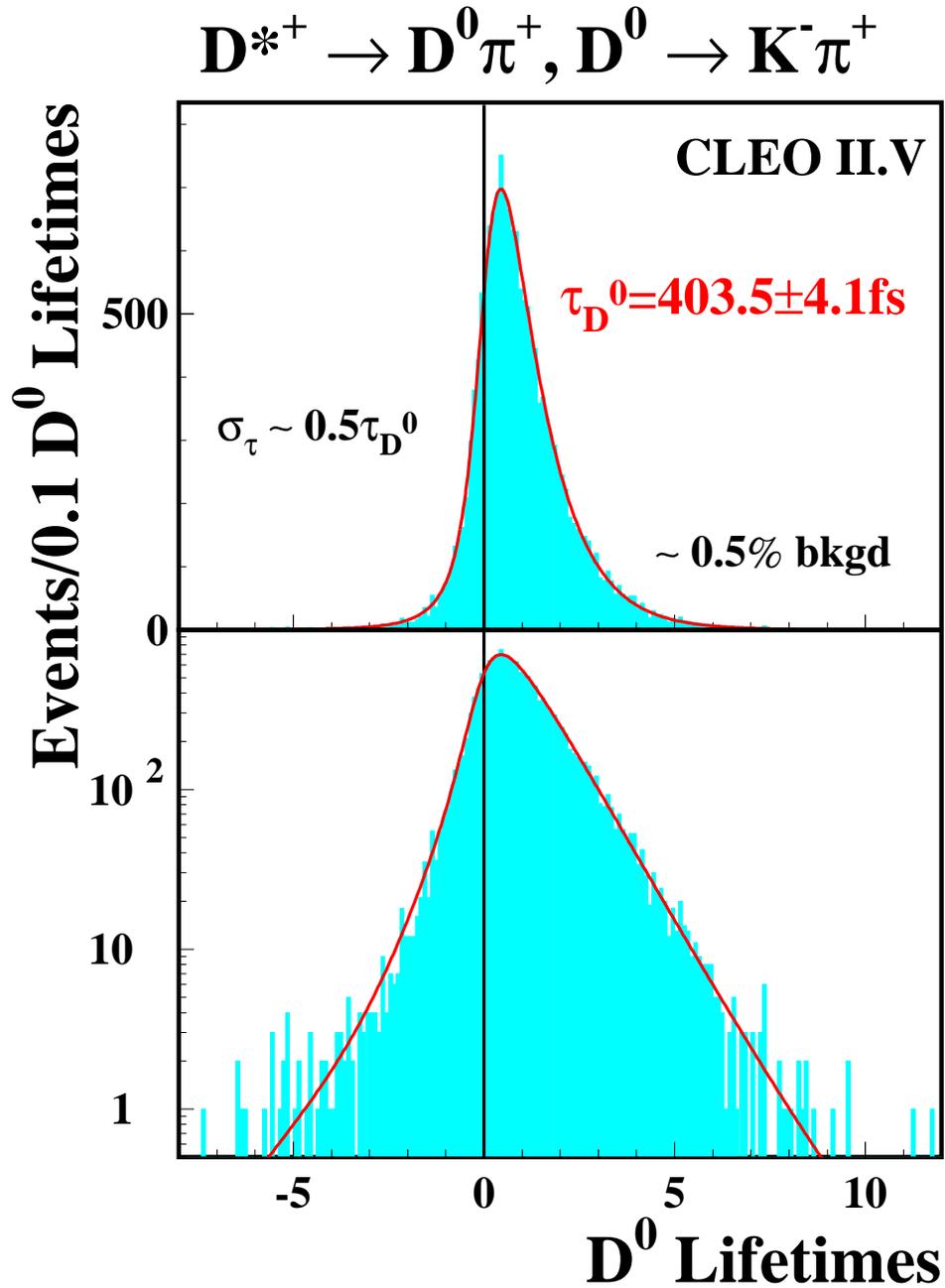,width=5.75in}
\end{center}
\caption[Distribution in $t$ for $\DZ\!\to\!K^-\pi^+$.]
{Distribution in $t$ for \hbox{right-sign} $\DZ\!\to\!K^-\pi^+$.
The histogram shows the data, and those data are the same for the
upper and lower plots; the only difference is the vertical scale, linear
above, logarithmic below.  The smooth curve is the fit, which results
in the mean life shown of $\tau_{\DZ}=403.5\pm4.1\,$fs, where
the error is statistical alone.  We have not assessed the systematic
error on the fit to $\tau_{\DZ}$, 
so, it should not be construed as a new result
on $\tau_{\DZ}$.}
\label{fig:lrs}
\end{figure}

In the sample of \hbox{right-sign}, $\DZB\!\to\!K^+\pi^-$ decays, $14199\pm120$
signal events pass the requirements for the reconstruction of $t$, out of
the initial sample of $16126\pm126$ used for the 
measurement of $R_{\rm ws}$.  The distribution in $t$ of the \hbox{right-sign}
events is shown in Fig.~\ref{fig:lrs}.  We fit that distribution in the
manner described in our measurement of mean charm lifetimes\cite{dlife},
and find, in units of the recent world average\cite{RPP98} mean $\DZ$
life, $0.972\pm0.014$, where no systematic error for this measurement
is assessed.  Since we have not assessed a systematic
error, this result should not be construed as a new result on $\tau_{\DZ}$.
However, possible systematic errors due to the reconstruction and 
fitting technique
are limited by this result from the \hbox{right-sign} data.  There
are far fewer events in the \hbox{wrong-sign} data,
and so this class of systematic errors is
negligible compared to the statistical error of the fit to the 
\hbox{wrong-sign} data.

The fit to the \hbox{right-sign} data determines the resolution
function that we use for the fit of the \hbox{wrong-sign} data.  We also
use the mean $\DZ$ lifetime from the fit to the \hbox{right-sign} data
as the central value for various charm backgrounds that are present
in the \hbox{wrong-sign} data.

\begin{figure}[htpb]
\begin{center}
\epsfig{figure=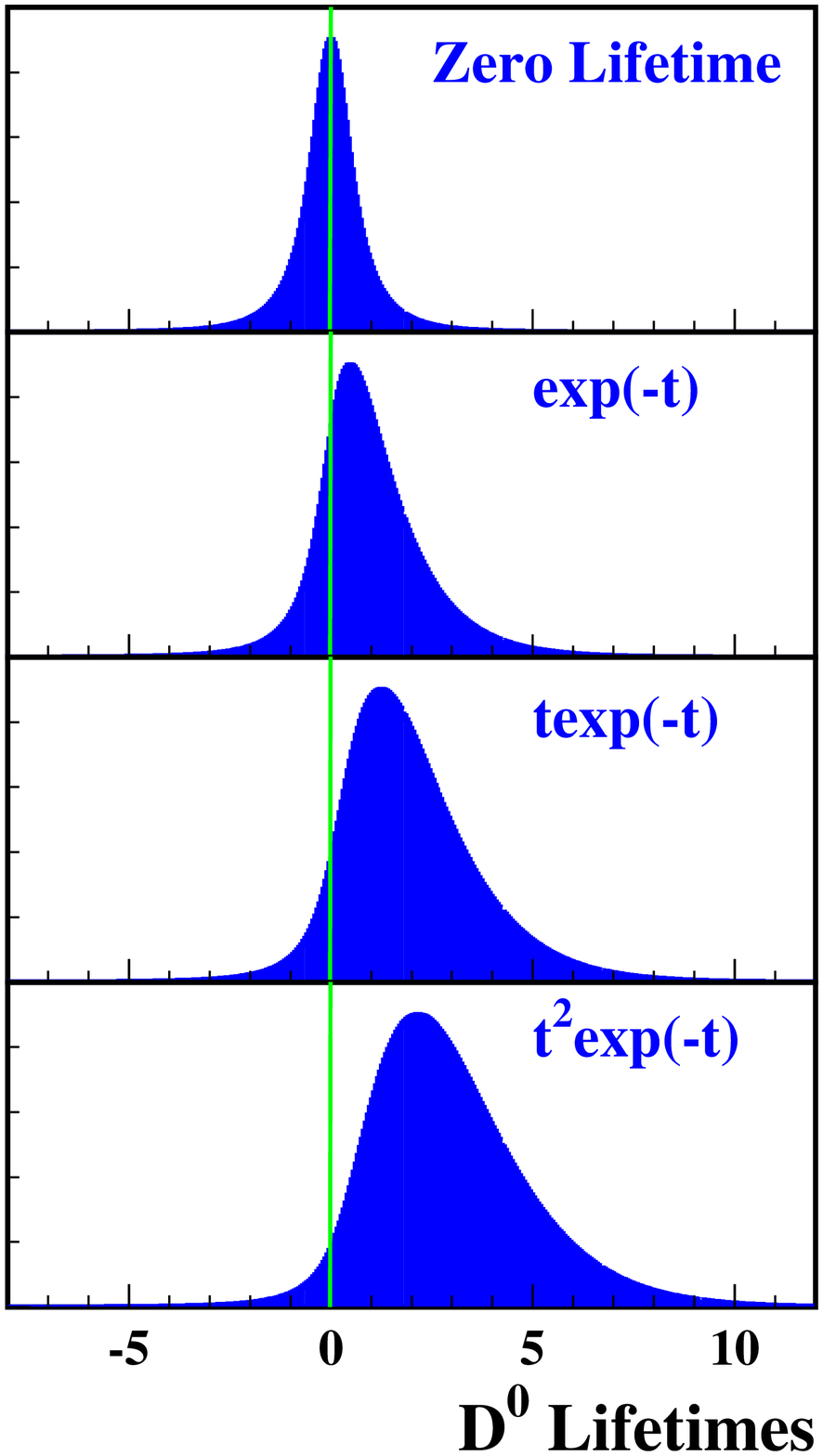,width=5.75in}
\end{center}
\caption[The Functional Forms of Decay and Mixing.]
{The functional forms of decay and mixing, as a function of
the proper time $t$, relative to the mean $\DZ$ lifetime.
The top plot shows our resolution function, as determined
from the fit in Fig.~\ref{fig:lrs}, which can
be approximately  characterized by a gaussian with 
resolution $\sigma_t\!\approx\!1/2$.
The subsequent three plots show this resolution function convolved with
the three functional forms evident in Eq.~\ref{eq:rws}:
$e^{-t}$, $te^{-t}$, and $t^2e^{-t}$, which describe, respectively,
DCSD decay, interference between DCSD decay and mixing, and mixing.}
\label{fig:4stack}
\end{figure}

Our resolution function is displayed, as well
as  the effect of folding that
function with each of the three functional forms, $e^{-t}$,
$te^{-t}$, and $t^2e^{-t}$, that enter in the
time-dependent rate of \hbox{wrong-sign} decays from 
Equation~\ref{eq:rws}, in Fig.~\ref{fig:4stack}.  For the
forms $te^{-t}$ and $t^2e^{-t}$, the breadth of the distribution
is dominantly from the functional forms themselves; a narrowing
of the resolution function would not greatly help distinguish between
them.

\begin{figure}[htpb]
\begin{center}
\epsfig{figure=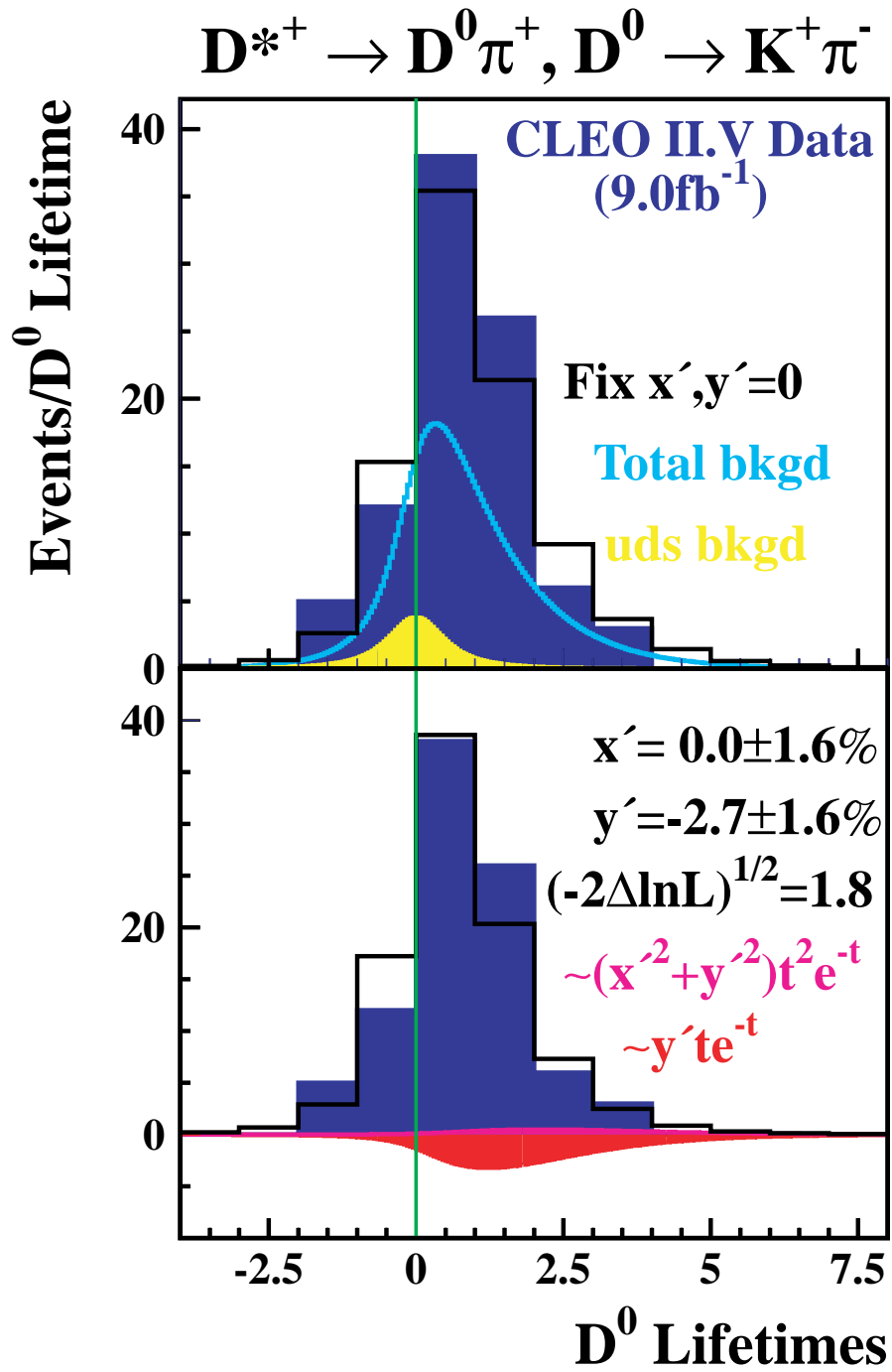,width=5.75in}
\end{center}
\caption[Distribution in $t$ for $\DZ\!\to\!K^+\pi^-$.]
{Distribution in $t$ for \hbox{wrong-sign} $\DZ\!\to\!K^+\pi^-$.
The same histogram of data is shown in both the upper and lower plots;
the differences are the fit results shown and overlayed.
The solid black lines are
the totals of all components of the fits.  The values of
$x^\prime$ and $y^\prime$, from Eqn.~\ref{eq:rws}, are fixed
to zero in the fit displayed in the upper plot. The best-fit shapes
from total and $uds$ backgrounds are shown as smooth curves.
The variables
$x^{\prime}$ and $y^{\prime}$ float freely in the lower fit,
and the best-fit values are shown. The likelihood decrement from the upper
to lower fit gives $\sqrt{-2\Delta\ln{\cal L}}=1.8\,\sigma$.  
The best-fit shapes of the two mixing
terms are shown, although only that from $te^{-t}$ is visible.}
\label{fig:lws}
\end{figure}

In the sample of \hbox{wrong-sign}, $\DZ\!\to\!K^+\pi^-$ decays, 91
events pass the requirements for the reconstruction of $t$, out of
the initial sample of 107 events in the signal region described
in Table~I, and used for the 
measurement of $R_{\rm ws}$.  The distribution in $t$ of the 91
\hbox{wrong-sign}
events is shown in Fig.~\ref{fig:lws}.  We fit that distribution in the
manner distinct from that used in our measurement of 
mean charm lifetimes\cite{dlife}.

We estimate the composition of the 91 events by applying the same
requirement to the (renormalized) Monte Carlo sample used to fit
for $R_{\rm ws}$. A summary of the composition appears in Table~III.

We fit the distribution of \hbox{wrong-sign} candidates displayed
in Fig.~\ref{fig:lws} with a superposition of $r_{\rm ws}(t)$, from
Equation~\ref{eq:rws}; an exponential with the $\DZ$ lifetime for the
`$\pi^{\pm }+\DZ/\DZB$' background; a zero-lifetime component
for the $uds$ background; and second exponential to describe
the $c\overline{c}$ and PV backgrounds.  All distributions in $t$
are folded with our resolution function.  A Poisson likelihood,
based on the extended maximum likelihood technique\cite{barlow},
is computed, for bins that are $1/10$ of a $\DZ$ lifetime, which
is much smaller than $\sigma_t$.

The mean lifetime of the Monte Carlo simulated data,
after renormalization for the fit to the
$Q\!-\!M$ plane, agrees very well with the data, outside
of the signal region.  This agreement gives us confidence
in the use of the Monte Carlo simulated data to describe
the background in the signal region.  In the Monte Carlo
simulated data, the decay time distribution of the $c\overline{c}$
and PV backgrounds is well described with a single exponential,
folded with the resolution function.  We use the mean $\DZ$ lifetime
to describe that exponential, but we explicitly vary that lifetime
by $\pm15\%$ to assess a systematic error.

The information on the expected fractions of the various background
components, as it appears in Table~III, is incorporated as a
a separate constraint added to the likelihood.  The error on each
fraction is nominally $(\sigma^2_P+\sigma^2_F)^{1/2}$, except
for the signal component, where $\sigma_F$ is omitted.  The systematic
errors, $\sigma_S$ from Table~III, are included when the systematic
errors on the fit parameters are evaluated.

\begin{center}
\begin{minipage}[t]{5.5in}
TABLE~III. Event yields in the $Q-M$ signal region
for wrong-sign $\DZ\!\to\!K^+\pi^-$ candidates
that have had $t$ successfully reconstructed.
The distribution in $t$ itself is shown in Fig.~\ref{fig:lws}.  
The event yields in the second column
are taken from the same fit to the data 
described in Table~I,
but with the effects of successful $t$ reconstruction
taken into account by the study of our sample
Monte Carlo simulated data, which corresponds to an
integrated luminosity of 90~fb$^{-1}$.
The statistical errors on the number of events, in the second
column, result from the fit to
the whole $Q\!-\!M$ plane; since the signal populates the
signal region with high efficiency, the errors on the 
$\DZ\!\to\!K^+\pi^-$ signal are the full Poisson errors on the
mean yield of events.  The statistical errors
on the background components from the fit
are relatively small compared to the Poisson
error that corresponds to the mean event yield;
a far greater number of background events populate 
the entire $Q\!-\!M$ plane than populate the signal region alone.
The third column gives the fraction of the wrong-sign
(WS) sample constituted by the signal and each category of background;
the fourth column gives the Poisson error $\sigma_P$, corresponding to
any one random sample drawn from a distribution with the
mean yield of events, on that fraction.
The fifth and sixth columns give the error $\sigma_F$
on each fraction from the fit
to the whole $Q\!-\!M$ plane, and then the error $\sigma_S$
from systematic effects, determined principally from varying
fit regions in the $Q\!-\!M$ plane.  For combining
errors, $\sigma_P$ and $\sigma_F$ are counted only once for the
signal.  The last column shows the mean life of the background
components, with our estimated systematic error on that mean life,
in units of the mean $\DZ$ lifetime.
For the signal $\DZ\!\to\!K^+\pi^-$, the mean life is unknown (unk.)
and can vary, according to Eqn.~\ref{eq:rws}, between 0.586 and 3.414.
For reference, the right-sign yield and measured mean life with
error is given in the last row.
\vspace{2ex}
\begin{center}
\begin{tabular}{cccccc|c}
   & \textbf{\#} & \textbf{\%}
 & \textbf{\%}
 & \textbf{\%} & \textbf{\%} & \\
\textbf{Component} & \textbf{Events} & \textbf{WS}
 & $\mathbf{\sigma_P}$
 & $\mathbf{\sigma_F}$ & $\mathbf{\sigma_S}$ & 
$\langle t \rangle$ \\ \hline
WS Data& $91$ & 100 &   &   &   & \\
$\DZ\!\to\!K^+\pi^-$ & $49.7\pm9.8$ & 54.6 & 10.8 & 10.8 & 8.7 & unk., .59-3.4 \\
\hline
\multicolumn{2}{l}{Backgrounds:} & & & & \\
$\pi^{\pm }+\DZ/\DZB$ & $19.6\pm1.4$ & 21.5 & 4.9 & 1.5 & 2.5 & $.972\pm0.014$ \\
$c\overline{c}$ & $10.4\pm1.1$ & 11.4 & 3.5 & 1.2 & 1.4 & $1.00\pm0.28$ \\
$uds$ & $5.1\pm0.6$ & 5.6 & 2.5 & 0.7 & 1.2 & $.004\pm0.015$ \\
PV & $6.2\pm0.7$ & 6.8 & 2.7 & 0.8 & 0.8 & $1.17\pm0.25$ \\ \hline
$\DZ\!\to\!K^-\pi^+$ & $14199$ & & & & & $.972\pm0.014$ \\ \hline
\end{tabular}
\end{center}
\end{minipage}
\end{center}

Six parameters are able to vary in the fit:
$x^{\prime}$, $y^{\prime}$, $a_D$ where $R_D=a_D^2$, all from
Equation~\ref{eq:rws}; and the fraction of each of the
$\pi^{\pm }+\DZ/\DZB$, combined $c\overline{c}+$PV, and $uds$ backgrounds.

The fit has been tested on 73 samples synthesized from the \hbox{right-sign}
candidates, and background from the sidebands in the \hbox{wrong-sign}
$Q\!-\!M$ plane.  Each of the synthesized samples contains 91 events, 
and they are composed to represent the wrong-sign sample described 
in Table~III.  The fits to the 73 synthetic samples show no
bias in fitted values of $x^\prime$, $y^\prime$, and $a_D$.
They also verify the ability of the fit to
estimate our statistical errors.

In the initial fit to the actual wrong-sign data,
$x^{\prime}$ and $y^{\prime}$ are constrained
to be zero, and this fit is shown in the upper part of
Fig.~\ref{fig:lws}; the confidence level of that fit is $84\%$,
indicating a good fit.

The mixing amplitudes $x^{\prime}$ and $y^{\prime}$ are then
allowed to freely vary, and the best fit values are shown in
both the lower portion of Fig.~\ref{fig:lws}, and in Table~IV.

\begin{center}
\begin{minipage}[t]{5.5in}
TABLE~IV. Results of the fit to the  
distribution of $\DZ\!\to\!K^+\pi^-$ in $t$.  Both the
distribution, and the fit, are shown in the lower portion
of Fig.~\ref{fig:lws}.
\vspace{2ex}
\begin{center}
\begin{tabular}{ccc}
\textbf{Parameter} & \textbf{Best Fit} & \textbf{95\% C.L.}\\ \hline
\vphantom{$A^{A^A}$}$R_D$      & $(0.52^{+0.11}_{-0.12}\pm0.08)\%$ & \\
$y^\prime$ & $(-2.7^{+1.5}_{-1.6}\pm0.2)\%$ & $-5.9\%<y^\prime<0.3\%$ \\ \hline
$x^\prime$ & $(0\pm1.6\pm0.2)\%$             & $|x|<3.2\%$ \\
$(1/2)x^{\prime2}$ &     & $<0.05\%$ \\ \hline
\end{tabular}
\end{center}
\end{minipage}
\end{center}

The fit improves slightly, by an amount corresponding to
$\sqrt{-2\Delta\ln{\cal L}}=1.8\,\sigma$, including systematic
effects, when mixing is allowed.  Our interpretation of this
change is that it represents a statistical fluctuation.

Therefore, our principal results concerning mixing are the
one-dimensional intervals, which correspond to a 95\% confidence level,
that are given in Table~IV.

\begin{figure}[htpb]
\begin{center}
\epsfig{figure=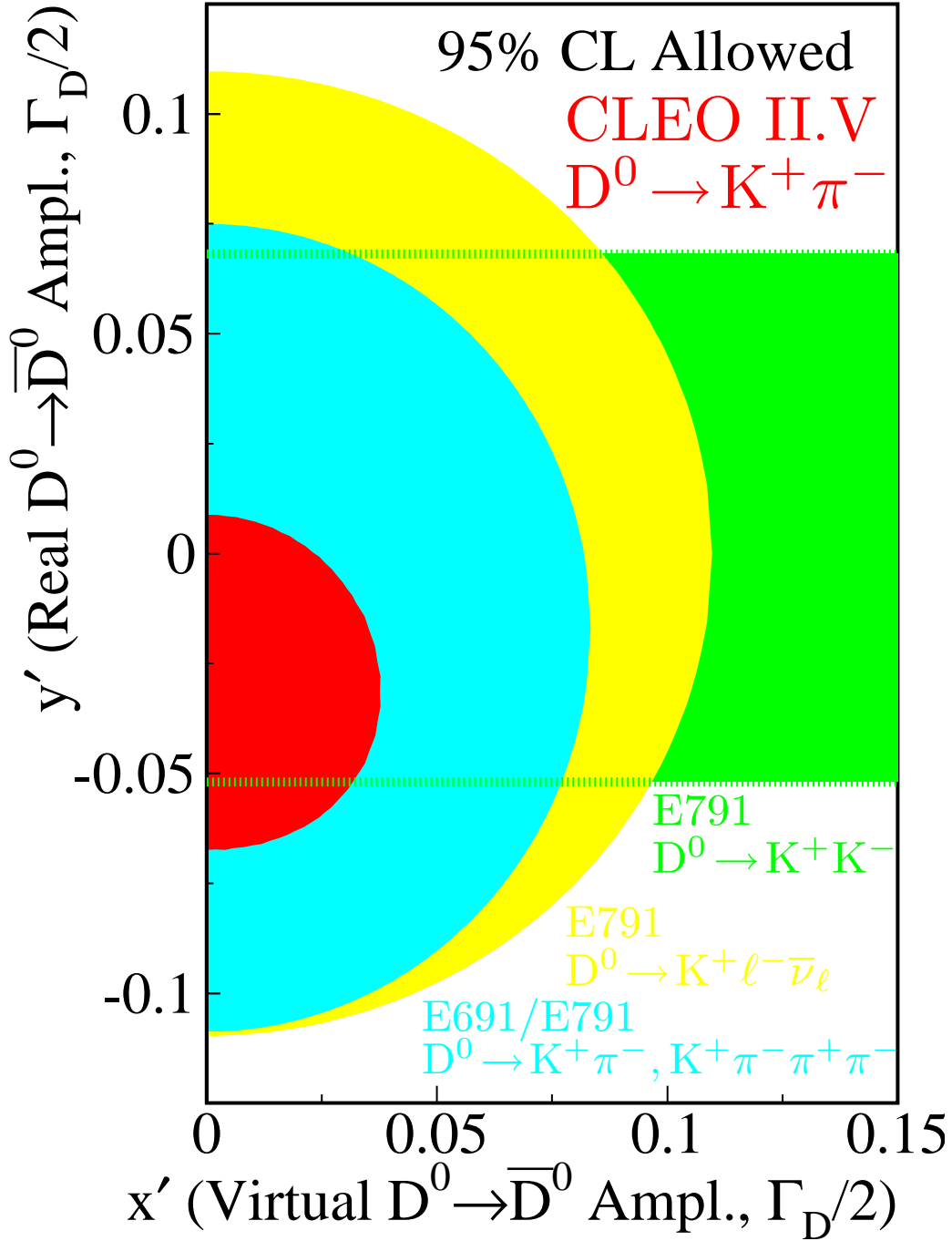,width=5.4in}
\end{center}
\caption[Limits in the $y^\prime$ v. $x^\prime$ Plane]
{Limits in the $y^\prime$ v. $x^\prime$ plane.  Our experiment
limits, at 95\% C.L., the true values of $x^\prime$ and
$y^\prime$ to occupy the dark (red) region near the origin.
Also shown are the similar zones from other recent experiments.
We assume $\delta=0$ to place the recent work of E791
that utilized $\DZ\!\to\!K^+K^-$;
a non-zero $\delta$ would
rotate the E791 confidence region clockwise about the origin
by an angle of $\delta$.}
\label{fig:xyl}
\end{figure}

Additionally, we evaluate a contour in the two-dimensional
plane of $y^{\prime}$ versus $x^{\prime}$, which at
95\% confidence level, contains the true value of $x^{\prime}$ and
$y^{\prime}$. To do so, we determine
the contour around our best fit values where the $-\ln{\cal L}$ increases
by 3.0~units.  The interior of the contour is shown, as the small
red region, near the origin of Fig.~\ref{fig:xyl}.
On the axes of $x^\prime$ and $y^\prime$, this contour
falls slightly outside the one-dimensional intervals listed in Table~IV,
as expected.

We have evaluated the allowed regions of other experiments
\cite{E691,E791,E791L,E791CP}, at 95\% C.L., and shown those
regions in Fig.~\ref{fig:xyl}.  In combining the
E691 and E791 studies of hadronic final states, we
make the most optimistic assumption, that leads to the
smallest allowed region, concerning treatment of the
term linear in $t$ in Equation~\ref{eq:rws}.  We do not
necessarily endorse that interpretation, and we note that
the E791 results utilizing $\DZ\!\to\!K^+\ell^-\overline{\nu}_{\ell}$
suffer no similar uncertainty in interpretation.

\begin{figure}[htpb]
\begin{center}
\epsfig{figure=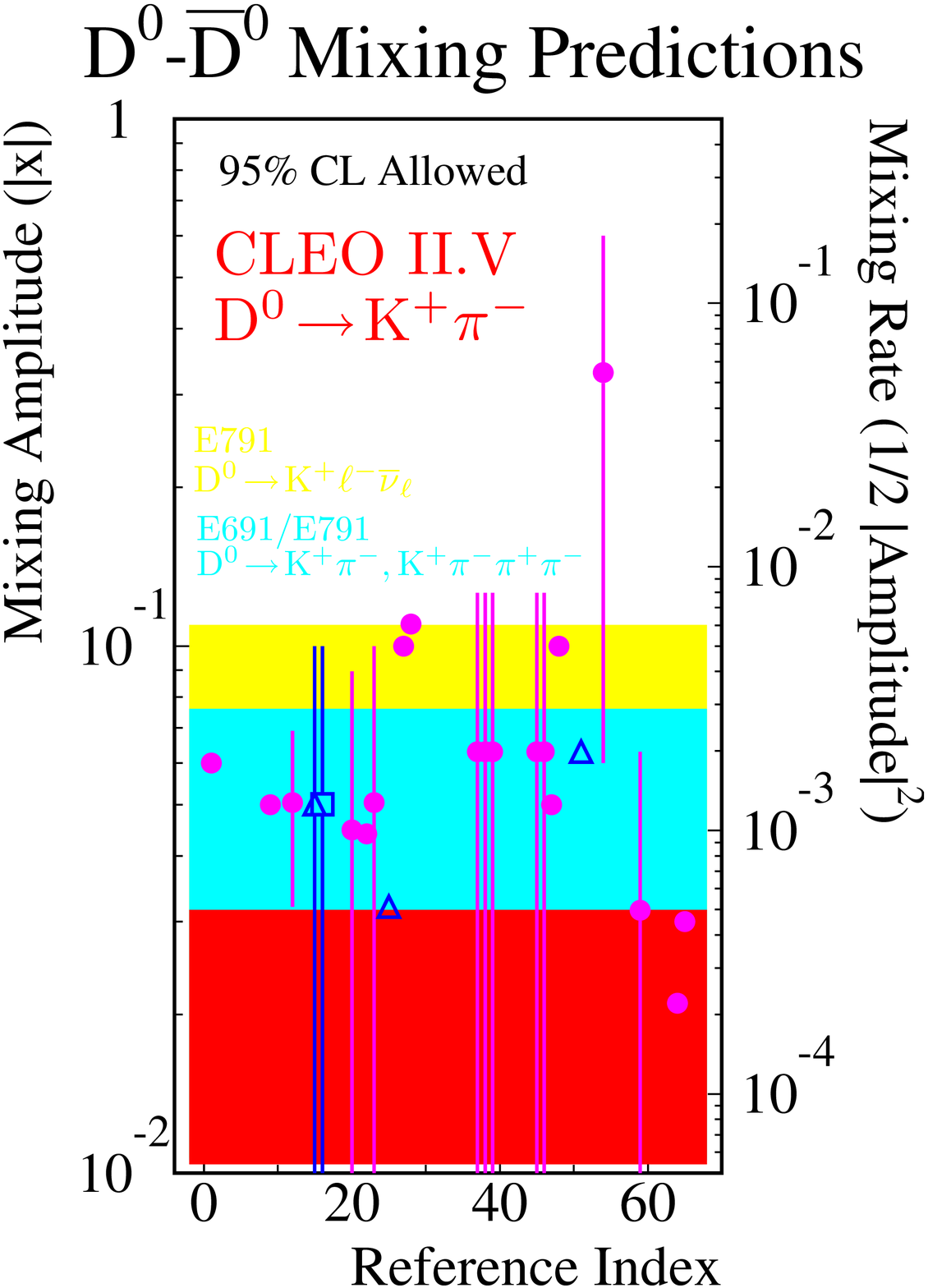,width=5.4in}
\end{center}
\caption[Predictions Impacted by this Work]
{Predictions impacted by this work.  The horizontal axis
is the Reference Index used in \protect\cite{hnncomp}, that
corresponds to the reference tabulated there.  The horizontal
region (red) nearest the bottom is the allowed region,
at 95\% CL, consistent with this work, in the plausible approximation
that $x^\prime\!=\!x$.  There are 18 predictions that have some
inconsistency with this work.}
\label{fig:pxln}
\end{figure}

Finally, if we assume that $\delta$ is small, which
is plausible\cite{wfsp,brpa}, then $x^\prime\!\approx\!x$,
and we can indicate the impact of our work in limiting predictions
of $\DZ\!-\!\DZB$ mixing from extensions to the Standard Model.
We plot our one-dimensional allowed region in $x^\prime$ in
Fig.~\ref{fig:pxln}.  Eighteen of the predictions tabulated
in a paper contributed to this symposium\cite{hnncomp} have
some inconsistency with our limit.  Among those predictions,
some authors have made common assumptions, however.

In conclusion, we conducted a study of the \hbox{wrong-sign}
process, $\DZ\!\to\!K^+\pi^-$, and conclusively established
its rate, relative to the \hbox{right-sign} process,
$\DZB\to\!K^+\pi^-$, as $R_{\rm ws}=(0.34\pm0.07\pm0.06)\%$.
By a study of that rate as a function
of the decay time of the $\DZ$, we distinguish the
rate of direct, doubly-Cabibbo-suppressed decay 
$\DZ\!\to\!K^+\pi^-$ relative to
$\DZB\!\to\!K^+\pi^-$, to be 
$R_D=(0.50^{+0.11}_{-0.12}\pm0.08)\%$.  
The amplitudes that describe
$\DZ\!-\!\DZB$ mixing, $x^\prime$ and $y^\prime$, are consistent
with zero.  The one-dimensional limits, at the 95\% C.L., that
we determine are $(1/2)x^{\prime2}<0.05\%$, and
$-5.9\%<y^\prime<0.3\%$.  The limit on $x^\prime$, combined
with the plausible assumption that the relative strong phase
$\delta\!=\!0$, has some inconsistency with a variety of
extensions to the Standard Model.
Limits in the two-dimensional
plane of $y^\prime$ versus $x^\prime$ are given in
Fig.~\ref{fig:xyl}.

All results described here are preliminary.

We gratefully acknowledge the effort of the CESR
staff in providing us with excelled luminosity
and running conditions.  We wish to acknowledge and
thank the technical staff who contributed to the success
of the CLEO II.V detector upgrade, including J.~Cherwinka and
J.~Dobbins (Cornell); M.~O'Neill (CRPP); M.~Haney (Illinois);
M.~Studer and B.~Wells (OSU); K.~Arndt, D.~Hale, and S.~Kyre (UCSB).
We appreciate contributions from G~Lutz and advice from A.~Schwarz.
This work was supported by the National Science Foundation,
The U.S. Department of Energy, Research Corporation, the
Natural Sciences and Engineering Research Council of Canada,
the A.~P.~Sloan Foundation, the Swiss National Science Foundation,
and the Alexander von Humboldt Stiftung.

\end{document}